\documentclass{article}
\usepackage{mathtools}
\usepackage{amsmath,yhmath}
\usepackage{bm}
\usepackage{graphicx}
\usepackage{caption}
\usepackage{subcaption}
\usepackage{soul}
\usepackage{amssymb}
\usepackage[dvipsnames]{xcolor}

\newcommand{\ii}{\mathrm{i}}
\newcommand{\dd}{\mathrm{d}}
\newcommand{\ee}{\mathrm{e}}

\newcommand{\rmvec}[1]{\mathrm{\mathbf{#1}}}

\newcommand{\pdv}[2]{\frac{\partial #1}{\partial #2}}

\usepackage{color}

\title{Switching via Wave Interaction in Topological Photonic Lattices} 

\author{%%%% Author details
Mark J. Ablowitz$^{1}$, Justin T. Cole$^{2}$ and S. D. Nixon$^{1}$}

\begin{document}

\maketitle

\begin{abstract}
A honeycomb Floquet lattice with helically rotating waveguides and an interface separating two counter-propagating subdomains is analyzed. Two topologically protected localized waves propagate unidirectionally along the interface.  %along the interface that exhibit unidirectional propagation. 
Switching can occur when these interface modes reach the edge of the lattice and %when 
the light splits into waves traveling in two opposite directions.
%and then %to travel along between the edge of the clockwise or counterclockwise sections. %of the lattice. 
The incoming mode, traveling along the interface, can be routed %directed to travel 
entirely or partially %partly %engineered %designed  to travel entirely 
along either lattice edge with the switching direction based on a suitable mixing %phase mixing 
of the interface modes. 
\end{abstract}

\section{Introduction}

%% BASICS
%Photonic 
Topological insulators are structures that rely on broken symmetry (time-reversal, inversion, etc.) to create non-trivial spectral topologies that allow the material to behave as an insulator on its interior, yet conductor along its boundary. %Interest in further 
% I REARRANGED THIS PARAGRAPH TO FOLLOW CLOSELY TO THE HISTORICAL PROGRESSION: CONDENSED MATTER PHYISCS FIRST, THEN OPTICS.
The study of topological insulators originates from the field of condensed matter physics and the \textit{Quantum Hall Effect} \cite{Klitzing1980, Tsui1982}. However, in general, that same phenomenon can be realized in carefully prepared periodic systems \cite{Haldane1988}, and specifically adapted to optical settings \cite{Haldane2008} with electromagnetic waves taking the place of electrons.
Connections between topology and optics as well as possible application for the edge modes conduct along the boundary has spurred considerable research in the field of topological optics \cite{RechtsmanReview, JoannopoulosReview}. A particularly relevant system consists of a photonic %optical
lattice with helically varying waveguides along the direction of propagation \cite{Rechtsman2013}. The helical waveguides mimic the effect of a magnetic field and break the time-reversal symmetry of the system.
% Motivated by \cite{Haldane2008}, topological edge waves were theoretically predicted and observed in magneto-optical systems \cite{MO2008,MO2009}. A photonic optical lattice with helically varying waveguides along the direction of propagation was introduced in \cite{Rechtsman2013}. The helical waveguides mimic the effect of a magnetic field and break the time-reversal symmetry of the system.

%% EDGES

 % This mechanism was used to realize a photonic topological insulator experimentally and demonstrate the existence of topologically protected edge states \cite{Rechtsman2013}. 
A key feature driving interest in topological insulator has been the existence of topologically protected edge states that propagate unidirectionally. The existence of such edge states is due to the bulk-edge correspondence \cite{Hatsugai1993} that relates topological invariants %the topological properties 
to the existence of chiral boundary modes %behavior on the boundary 
\cite{BulkEdge2000}. 
%\jtc{WHY SPECIFICALLY FOCUS ON MO HERE?} 
These edge states can be physically realized in magneto-optical systems as outlined in \cite{Raghu2008} and experimentally realized in \cite{WangChong2009}. Since the edge states are back-scattering immune 
\cite{WangChong2008}, they are not only robust to defects in the lattice \cite{Hafezi2013, Hafezi2013B} but travel around sharp corners in the lattice \cite{Ma2015,YiPingMa2015}. Linear and nonlinear edge states have been investigated analytically using tight-binding approximations for honeycomb \cite{Curtis2014, Cole2017}, Lieb and kagome \cite{Cole2019} lattices with helical waveguides  in the paraxial (Schr\"{o}dinger) equation. %\jtc{THE PREVIOUS THREE SENTENCES CONTAIN  DIFFERENT PHYSICAL SYSTEMS. THESE ANALYTICAL RESULTS DON'T DESCRIBE MOST OF THEM. PERHAPS A RE-ARRANGEMENT IS APPROPRIATE} 
The topological properties of the lattice are calculated from the Berry curvature and characterized by the Chern number (an integer), whose sign determines which direction edge modes propagate along the boundary. 

%% Domain Walls

In this article, we present an optical switching method in a honeycomb lattice with helically rotating lattice sites. A vertical interface separates two counter-propagating subdomains: one counter-clockwise and the other clockwise. There are several methods used %employed 
in the creation of helically driven waveguides including writing directly onto fused silica glass \cite{Li2017, RechtsmanQVH}, optical induction in a nonlinear crystal \cite{OpticalInduction2011}, and multi-beam interference \cite{MultiBeam2011}. In this last method, control of the helical pitch may be achieved by changing the phase of the lateral beam to create the sort of interface we purpose to study here \cite{InterfaceZhigang, Shi2019}.  Other experimental designs that support interface modes on a domain wall have been closely tied to the quantum valley hall effect \cite{DWTheory1, DWTheory2}. In silicone based graphene, an interface can be created between regions of distinct buckled geometry \cite{SiliconeVQH2014,Silicene2015, Silicene2019} or all-dielectric photonic crystal \cite{MaVQH2016}. In \cite{RechtsmanQVH}, a detuning of the refractive indices for two honeycomb sublattices creates a narrow strip with differing Chern number. In this system, the interface modes do not posses a nontrivial Chern number and thus not immune to back scattering.

%% SWITCHES

In this work, we find interface modes that are localized around the interface and decay exponentially into the lattice, perpendicular to the interface. Just like edge modes, these interface modes exhibit a unidirectional propagation along the interface. %when modulated by an envelope function. 
A switching phenomena occurs when the interface modes reach an interface terminus and the light splits and travels along the left and/or right lattice. Through a judiciously chosen  linear combination %suitable choice 
of the interface states, it is possible to control the flow of light. The idea of creating topological switches is well-known and examples have been made in a wide range of settings including electromagnetics \cite{ElectromagneticSwitch}, phononics \cite{PhononicSwitch}, nonlinear optics \cite{NLOpticsSwitch}, and more. However, previous methods have relied on the manipulation of the lattice itself to change the topological properties of 
the lattice. A recent magneto-optical lattice has experimentally observed mode switching that can adjusted through an external magnetic field or through the modal frequency  \cite{MaSwitching2023}. 

In our approach, the lattice is fixed and the switching is achieved by adjusting the relative magnitudes and phases between a linear superposition of interface modes. In effect, the interfaces "remembers" it's switching instructions, realizing a nonlocal switching mechanism. While we restrict ourselves to the honeycomb lattice other lattice structures that exhibit non-trivial Chern topology can also be used \cite{Nixon2022}. In the future, nonlinear effects will be considered.

\subsection{Governing Wave Equation}
\label{Sec: PotentialNLS}

Small amplitude electromagnetic waves propagating through a waveguide array in the paraxial regime satisfy the Schr\"{o}dinger equation with an external potential
\begin{equation}
   \ii \pdv{\Psi}{z} + \nabla^2\Psi + V(\rmvec{r}, z) \Psi = 0 ,
   \label{Eq: NLS}
\end{equation}
written here in normalized form, where $\Psi$ is the envelope of the electric field, $z$ is the propagation distance, $\rmvec{r} = (x,y)$ is the transverse plane, and the potential $V(\rmvec{r}, z)$ is the index of refraction for a lattice with helically rotating lattice sites.

\begin{figure}[ht]
    \center
        \includegraphics[width=0.6\textwidth]{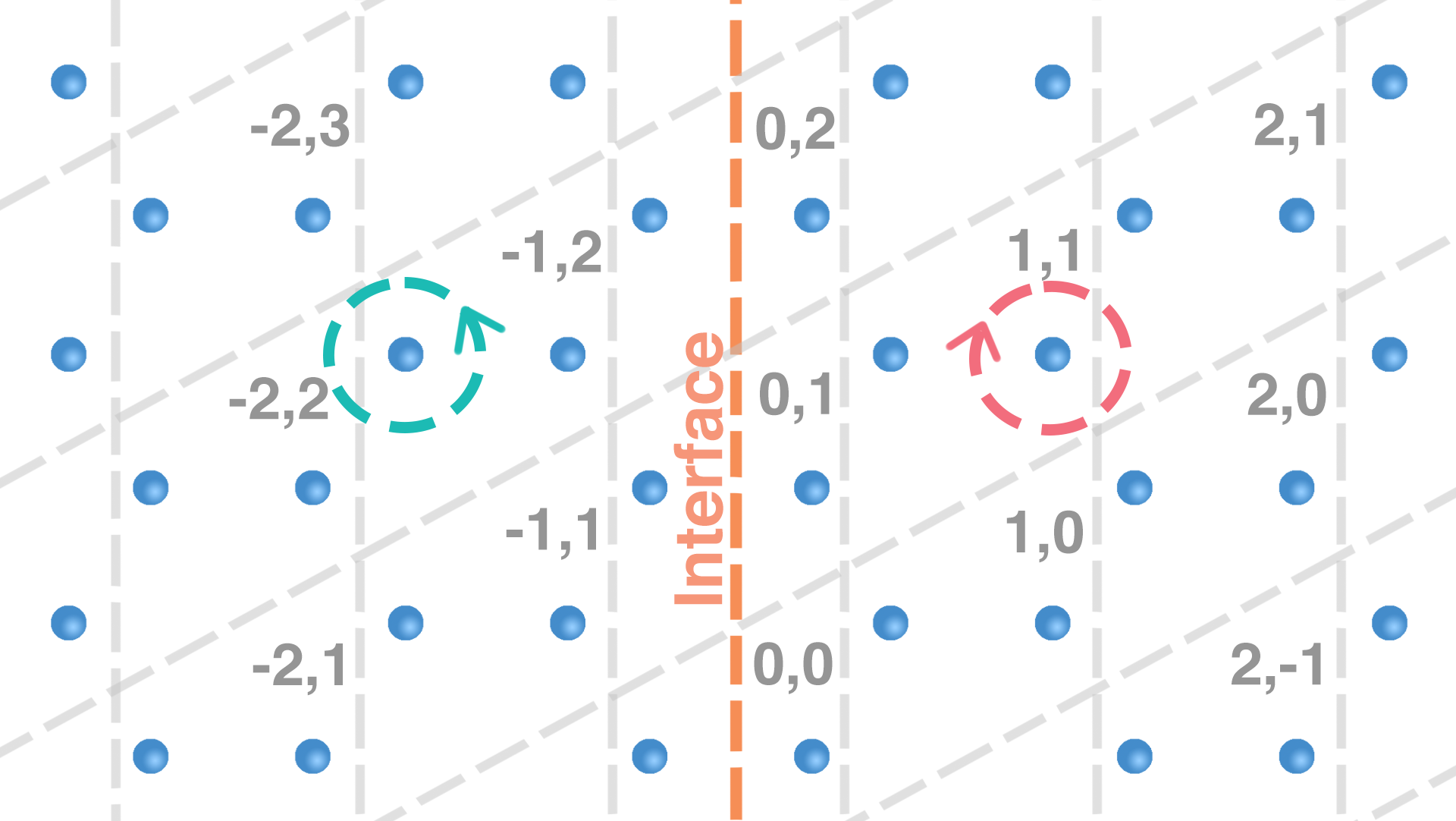}\includegraphics[width=0.3\textwidth]{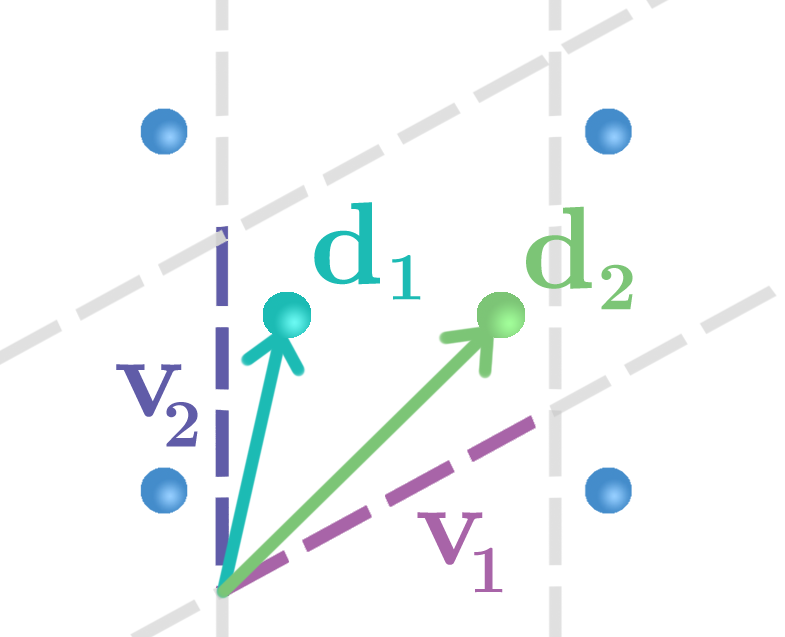}
        \caption{Honeycomb lattice with an interface between counter rotating lattice sections. The $n,m$ index gives the lattice cell with bottom left corner at $\rmvec{r}_{nm} = n \rmvec{v}_1 + m\rmvec{v}_2$. The ${\bf d}_1$ (${\bf d}_2$) vector gives the starting location for the $a$ ($b$) lattice site.}
            \label{Fig: HCLattice}
\end{figure}

We consider the case of a honeycomb lattice with a vertical interface as depicted in Figure \ref{Fig: HCLattice}. The undriven (stationary) lattice is doubly periodic in the transverse plane and can be represented by a tiling of parallelograms, lattice cells, with sides defined by the vectors $\rmvec{v}_1 = (3/2,\sqrt{3}/2)$ and $\rmvec{v}_2 =(0,\sqrt{3})$. The position of individual lattice sites is given by
\begin{equation}
    \rmvec{r}_{nm}^{\ell} = \rmvec{d}_{\ell} + n \rmvec{v}_1 + m \rmvec{v}_2 , ~~~~ \ell = 1,2
\end{equation}
where the vectors $\rmvec{d}_1 = \left(\frac{1}{4},\frac{3\sqrt{3}}{4} \right)$ and $\rmvec{d}_2 = \left(\frac{5}{4},\frac{3\sqrt{3}}{4} \right)$ give the positions of the $a$ and $b$ sublattices relative to the corner of the lattice cell at $\rmvec{r}_{nm}$, as illustrated in right insert of Figure \ref{Fig: HCLattice}. Alternatively, the lattice can be expressed in terms of the %connection 
vectors $\rmvec{w}_1 = (1,0)$, $\rmvec{w}_2 = (-1/2 \sqrt{3}/2)$, and 
$\rmvec{w}_3 = (-1/2, -\sqrt{3}/2)$ that point from an $a$ sublattice site to the nearest neighbors on the $b$ sublattice. We split the lattice into a left sublattice for $n<0$ and the $a$ site of $n=0$, and a right sublattice for $n>0$ and the $b$ site of $n=0$.

The interface is created by %from 
a difference in the orientation of helical rotations %for lattice sites 
on the left and right lattice, represented by the driving functions $\rmvec{h}_L(z)$ and $\rmvec{h}_R(z)$, respectively. The entire lattice is now defined by
\begin{align}
        V(\rmvec{r},z) = &\displaystyle\sum_{m=-\infty}^{\infty}\Bigg( \widetilde{V}(\rmvec{r} - \rmvec{r}_{0m}^{(1)}-\rmvec{h}_L(z)) +\sum_{n=-\infty}^{-1}\sum_{\ell = 1}^2 \widetilde{V}(\rmvec{r} - \rmvec{r}_{nm}^{\ell} -\rmvec{h}_L(z))\notag\\
        &\hspace{0.7cm}+ \widetilde{V}(\rmvec{r} - \rmvec{r}_{0m}^{(2)} -\rmvec{h}_R(z))+\sum_{n=1}^{\infty}\sum_{\ell = 1}^2 \widetilde{V}(\rmvec{r} - \rmvec{r}_{nm}^{\ell}-\rmvec{h}_R(z))\Bigg) ,
        \label{Eq: Potential}
\end{align} 
where
\begin{equation}
    \widetilde{V}\big(\rmvec{r}) = V_0^2\ee^{-x^2-y^2}
\end{equation}
is the uniform shape of the lattice sites, and $V_0\gg1$ is a large constant proportional to refractive index contrast of the waveguides. The interface runs parallel to $\rmvec{v}_2$ and in between the $a$ and $b$ lattice sites at $n=0$. We take driving functions of the form 
\begin{subequations}
    \begin{align}
        \rmvec{h}_L(z) &= \left( \kappa \cos(\omega z +\tau_L), \kappa \sin(\omega z +\tau_L) \right)\\
        \rmvec{h}_R(z) &= \left( \kappa \cos(-\omega z +\tau_R), \kappa \sin(-\omega z +\tau_R) \right),
    \end{align}
\end{subequations}
for $\omega > 0$, so that the left lattice rotates counterclockwise and the right lattice rotates clockwise.

For a deep lattice, we seek solutions of the form 
\begin{align}
   \label{Eq: Ansatz}
    \psi = \sum_{(n,m) \in  \mathbb{Z}^2} &\Bigg[a_{nm}(z) \phi\big(\rmvec{r}-\rmvec{r}_{nm}^{(1)} - \rmvec{h}_{n}(z) \big) \\ & + b_{nm}(z) \phi\big(\rmvec{r}-\rmvec{r}_{nm}^{(2)} - \rmvec{h}_{n+1}(z) \big)\Bigg] \ee^{-\ii \mu_0 z} \notag
\end{align}
where $\rmvec{h}_{n}(z) = \rmvec{h}_L(z)$ for $n\leq0$ and $\rmvec{h}_{n}(z) = \rmvec{h}_R(z)$ for $n>0$, $\mu_0$ is the dispersion relation, and $\phi(\rmvec{r})$, the orbital function, satisfies the equation
\begin{equation}
    \mu_0 \phi + \nabla^2 \phi + \widetilde{V}(\rmvec{r}) \phi  = 0.
    \label{Eq: OrbitalEquation}
\end{equation}
As detailed in Appendix \ref{APP: Details}, under a deep lattice assumption, $V_0 \gg 1$, the solution dynamics are described by a discrete system of equations for amplitude functions $a_{nm}$ and $b_{nm}$.

For the left and right lattices, the displacement between lattice sites remains fixed even as the entire lattice rotates, and it is helpful to express the lattice in terms of the connection vectors $\rmvec{w}_1$, $\rmvec{w}_2$, $\rmvec{w}_3$, defined above. On both sides of the interface, the governing system of equations is 
\begin{subequations}
\begin{align}
    \ii \frac{\dd}{\dd z} a_{nm}(z)&=q_j(z,\rmvec{w}_1) b_{nm} + q_j(z,\rmvec{w}_2)b_{n-1,m+1}+q_j(z,\rmvec{w}_3)b_{n-1,m} \label{Eq: LeftRightNM1}\\
        \ii \frac{\dd}{\dd z} b_{nm}(z)&=q_j^*(z,\rmvec{w}_1) a_{nm} +q_j^*(z,\rmvec{w}_2)a_{n+1,m-1}+q^*_j(z,\rmvec{w}_3)a_{n+1,m}
        \label{Eq: LeftRightNM2}
\end{align}
where the coefficients are given by 
\begin{align}
 q_j(z,\rmvec{w}) = \Big(q_0- \ii s_0 {\bf h}_j'(z) \cdot \rmvec{w} \Big)\ee^{-\sqrt{\mu} \Vert \rmvec{w} \Vert},
 \label{Eq: CoefFunc}
\end{align}
with $j=L$ on the left lattice, $n<0$, and $j=R$ on the right lattice, $n>0$. Here $*$ stands for complex conjugation.d
Along the interface, at $n = 0$, the displacement between the lattice sites across the boundary, from $a$ site to $b$ site, is $\rmvec{w}_1 + \Delta \rmvec{h}(z)$ where $\Delta \rmvec{h}(z) = \rmvec{h}_R(z) - \rmvec{h}_L(z)$ is the stretching of the lattice at the interface due to the counter rotation (counter driving) %propagation 
of the the two sides. %helical rotation. 
The coefficient function (\ref{Eq: DiscreteLatticeEquations}) %(\ref{Eq: CoefFunc})
 is tied to the location of the nearest neighbor lattice sites, so lattice sites near the interface will depend on both the left and right driving functions. The governing equations near the interface are given by
\begin{align}
    \ii \frac{\dd}{\dd z} a_{0,m}(z)&=q_R\big(z,\rmvec{w}_1+\Delta \rmvec{h}\big) b_{0,m} + q_L(z,\rmvec{w}_2)b_{-1,m+1}+q_L(z,\rmvec{w}_3)b_{-1,m}
         \label{Eq: InterfaceNM1}\\
        \ii \frac{\dd}{\dd z} b_{0,m}(z)&=q_L^*\big(z,\rmvec{w}_1+\Delta \rmvec{h}\big) a_{0,m} +q_R^*(z,\rmvec{w}_2)a_{1,m-1}+q^*_R(z,\rmvec{w}_3)a_{1,m}
        \label{Eq: InterfaceNM2}
\end{align}
\label{Eq: DiscreteLatticeEquations}
\end{subequations}

\section{Discrete-Continuous Model and Interface Modes}

The lattice is divided into three different regions: left lattice, right lattice and interface. The lattice is assumed to be periodic along the direction %of 
parallel the interface, $\rmvec{v}_2$, so we can transform the model into the Fourier space to calculate the interface modes. We introduce the change of variables
\begin{subequations}
    \begin{align}
    \label{edge_soln1}
        a_{nm}(z) = \alpha_n(z, k) \ee^{\ii 2\pi \left(m + \frac{n}{2}\right) k}\\  \label{edge_soln2}
        b_{nm}(z) = \beta_n(z, k) \ee^{\ii 2\pi \left(m + \frac{n}{2}\right) k }
    \end{align}
\end{subequations}
where the extra $n$ term in the phase accounts for the angle of the lattice vector $\rmvec{v}_1$. Substituting this into equation (\ref{Eq: LeftRightNM1}) and (\ref{Eq: LeftRightNM2}) to arrive at equations for the left and right lattices
\begin{subequations}
\begin{align}
    \ii \frac{\dd}{\dd z} \alpha_n(z,k)&=q_j(z,\rmvec{w}_1) \beta_n + \Big(q_j(z,\rmvec{w}_2)\ee^{\ii  \pi k}+q_j(z,\rmvec{w}_3)\ee^{-\ii \pi k}\Big)\beta_{n-1}\\
        \ii \frac{\dd}{\dd z} \beta_n(z,k)&=q_j^*(z,\rmvec{w}_1) \alpha_n +\Big(q_j^*(z,\rmvec{w}_2)\ee^{-\ii\pi k}+q^*_j(z,\rmvec{w}_3)\ee^{\ii \pi k}\Big)\alpha_{n+1}
\end{align}
where $j=L$ on the left lattice for $n<0$, and $j=R$ on the right lattice for $n>0$. Substituting into the interface equations (\ref{Eq: InterfaceNM1}-\ref{Eq: InterfaceNM2}) at $n=0$ gives
\begin{align}
    \ii \frac{\dd}{\dd z} \alpha_0(z,k)&=q_R(z,\rmvec{w}_1+\Delta\rmvec{h}) \beta_0 +\Big(q_L(z,\rmvec{w}_2)\ee^{\ii  \pi k}+q_L(z,\rmvec{w}_3)\ee^{-\ii \pi k}\Big)\beta_{-1}\\
        \ii \frac{\dd}{\dd z} \beta_0(z,k)&=q_L^*(z,\rmvec{w}_1+\Delta\rmvec{h}) \alpha_0 +\Big(q_R^*(z,\rmvec{w}_2)\ee^{-\ii\pi k}+q^*_R(z,\rmvec{w}_3)\ee^{\ii \pi k}\Big)\alpha_{1}.
\end{align}
\label{Eq: LatticeNK}
\end{subequations}

\begin{figure}
    \centering
    \includegraphics[height = 5cm]{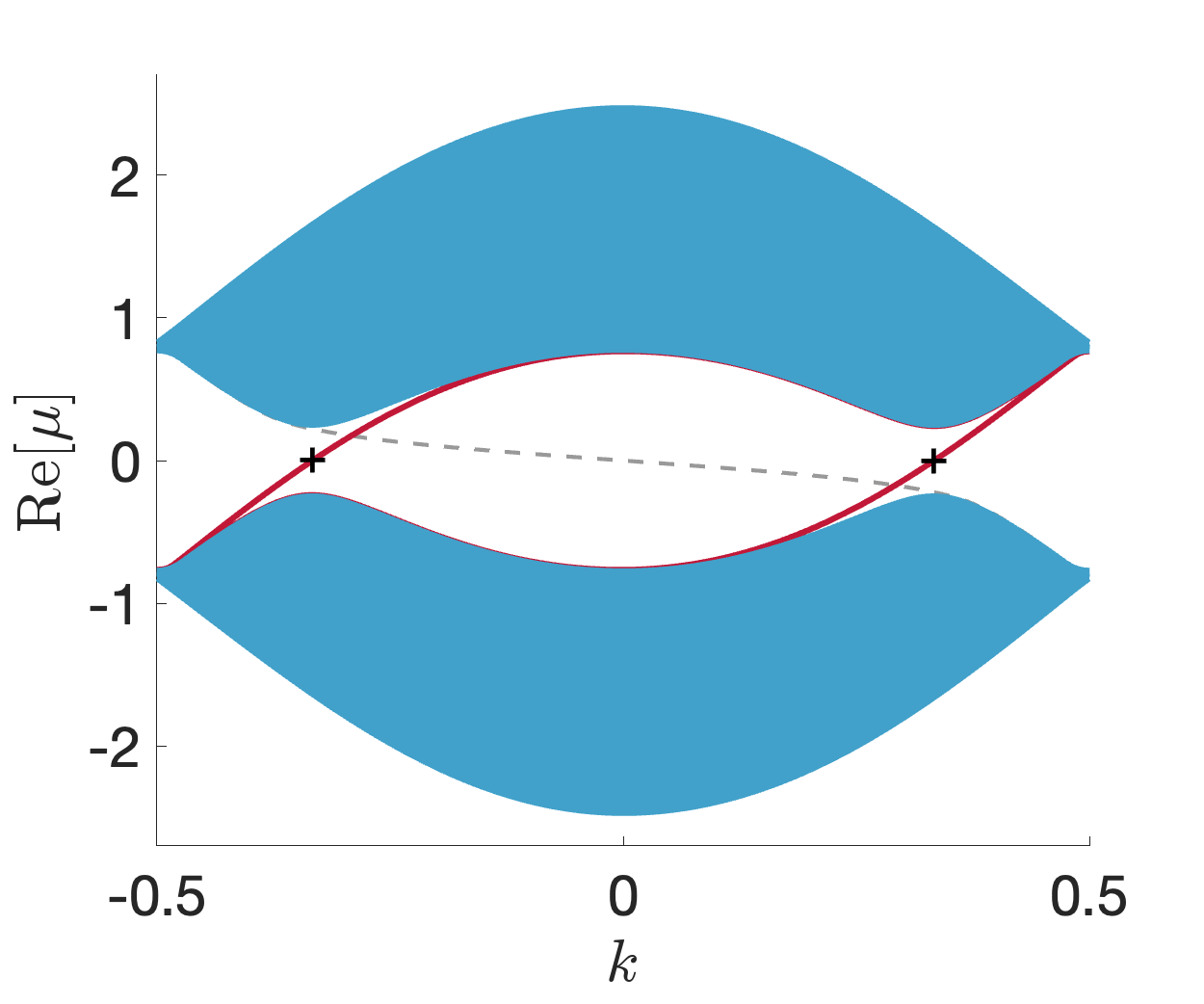}
    \includegraphics[height = 5cm]{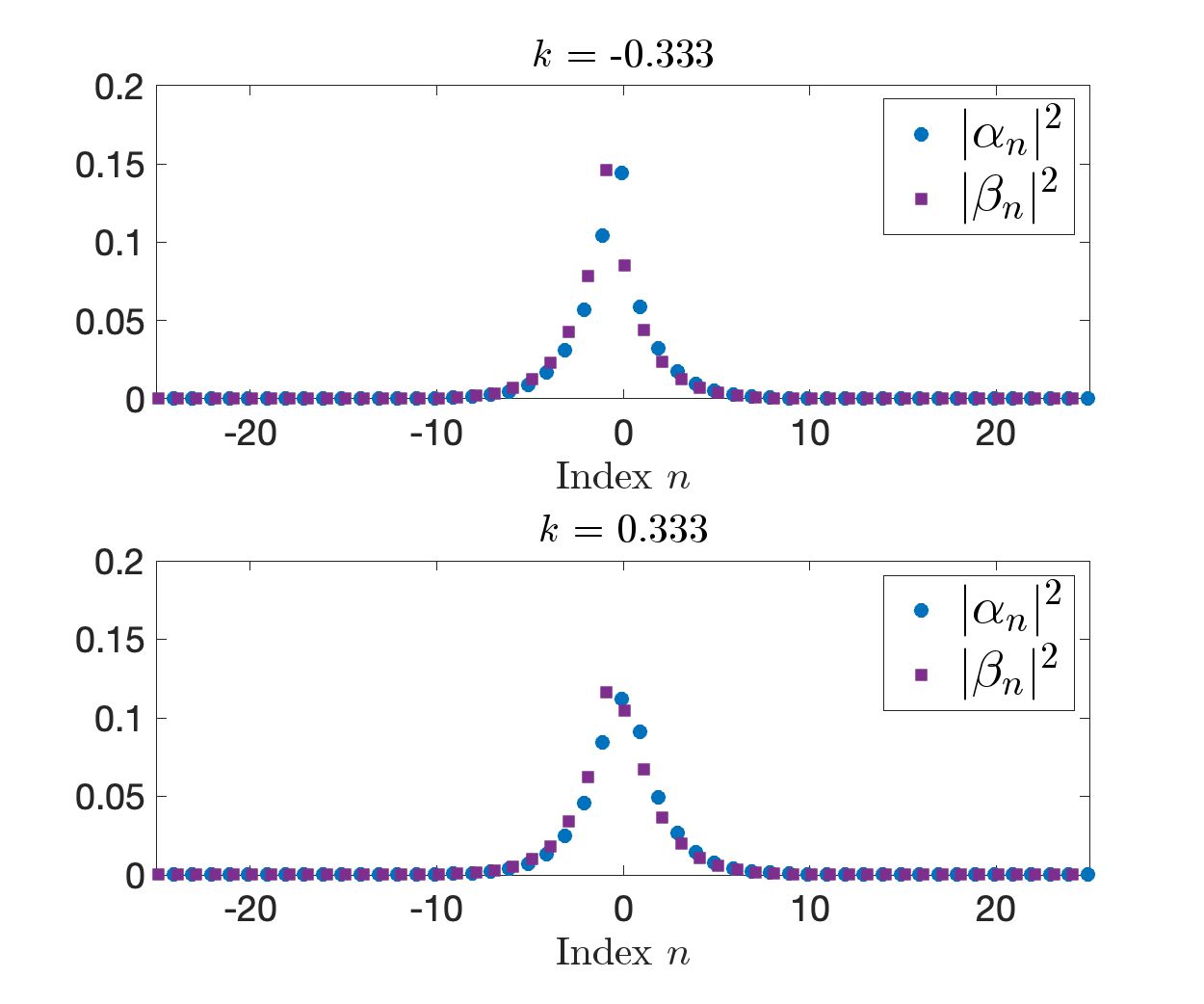}
    \caption{Spectrum and Floquet modes %eigenvectors 
    for the system (\ref{Eq: LatticeNK}) with $q_0 = 15$, $s_0 = 8$, $\kappa = 0.3$, $\omega = 2 \pi$, $\tau_L = \pi/2$ and $\tau_R = 3\pi/2$. (Left) Spectrum with the two families of interface modes that span the gap shown in red. Specific interface modes are marked by $+$.  (Right) The two interface modes found for $\mu = 0$. }
   \label{Fig: SpectrumModes}
\end{figure}

The infinite lattice (in $n$) is approximated by a wide lattice with boundary conditions $\alpha_n(k,z) = \beta_n(k,z) = 0 $ for $n<-50$ and $n>50$. We look for Floquet solutions $\alpha_n(z,k),\beta_n(z,k)$  proportional to $\exp{(i\mu z)}$ and solve %solving 
the resulting Floquet %eigenvalue 
problem for $\mu (k)$. The first two bands are given in Fig. \ref{Fig: SpectrumModes}.

The families of topological interface modes that span the gap %and 
are the main focus of the article. These families of solutions are supported by the counter rotating (mutually reinforcing) %propagation of the 
driving in the left and right sublattices and the change in topology across the interface. As we see in Figure \ref{Fig: SpectrumModes}, the topological interface modes come in pairs for given spectral parameter, $\mu$. This means for a chosen value of $\mu(k)$ in the gap, there is 
a two dimensional eigenspace of interface modes; this allows for a degree of freedom when constructing initial conditions to send along the interface. Within the continuous spectrum there there exist regions with a minor instability, $\mathrm{Im}[\mu]\neq 0$. However, these modes and do not appear in the direct numerics presented in figures \ref{Fig: Splitting} and \ref{Fig: Sensor} and do not interfere with the optical switching. 
%\jtc{AFTER OUR CONVERSATION EARLIER TODAY, CAN WE REMOVE THESE 2 PREVIOUS SENTENCES NOW AND REPLACE $Re[\mu]$ WITH JUST $\mu$ IN Fig 2?} 
The dashed 
 line represents the edge modes resulting from the numerical truncation of the domain and do not exist in the theoretical infinite lattice with an interface. Both the left and the right edge modes have the same eigenvalue.

\section{Optical Switching}

Consider an interface lattice where the lattice ends in an armchair boundary perpendicular to the interface, i.e., $a_{nm}=b_{nm}=0$ for $\frac{n}{2} + m>0$. Next, an initial carrier-envelope wave consisting of a topological interface mode travels up along the interface until it reaches the top armchair boundary, at which point the energy is found to split between %along 
the left and right subdomains. In this model there is negligible %insignificant 
energy loss during the spitting process.
There are two families of interface modes that span the gap (recall % see 
Figure \ref{Fig: SpectrumModes}), however there is only a single edge mode that propagates unidirectionally in either direction along the armchair boundary. By mixing the two interface modes, the combined pulse can be adjusted to split the pulse completely to the left, completely to the right, or evenly.

\begin{figure}
    \centering
    \includegraphics[height = 4.4cm]{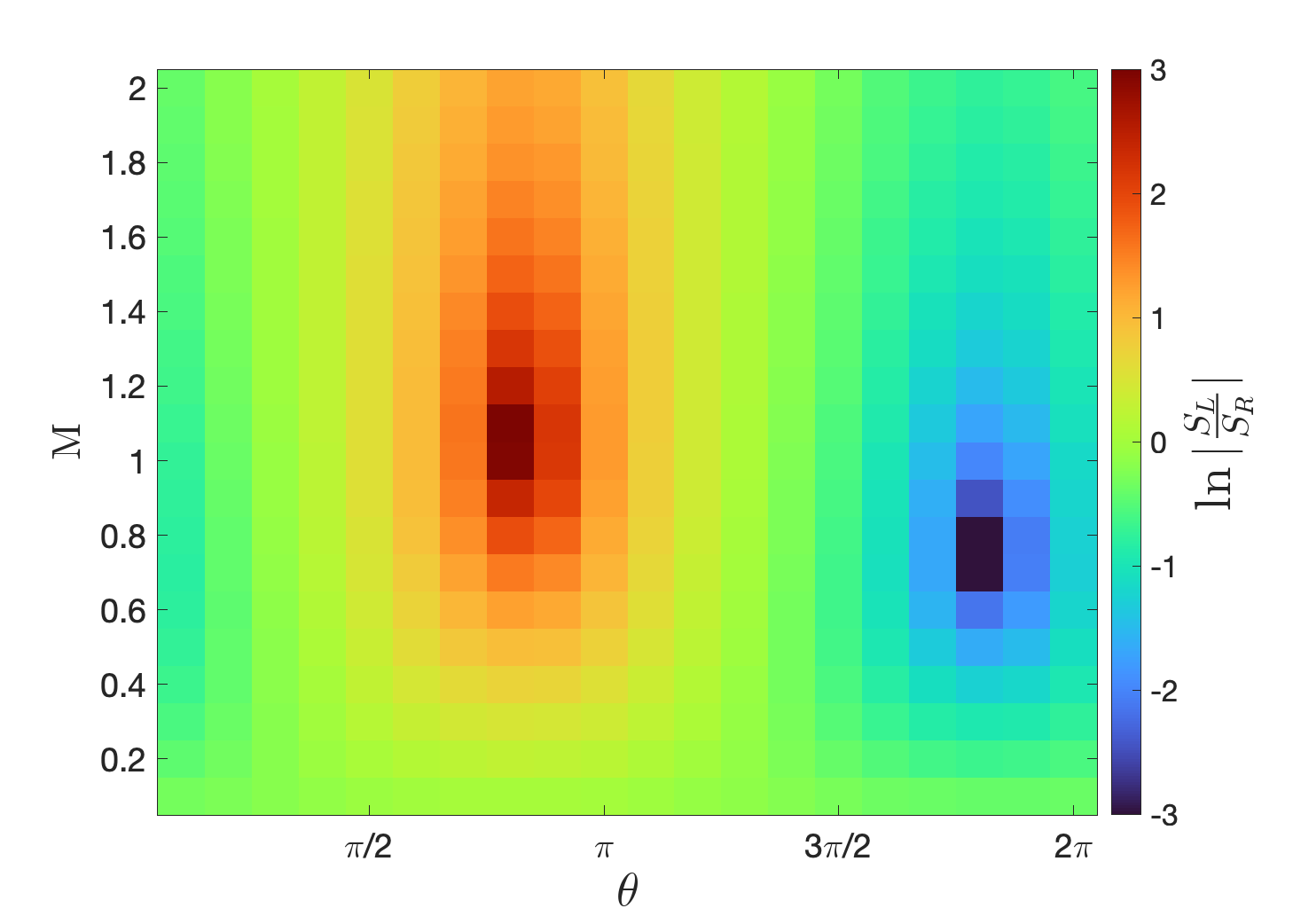}
    \includegraphics[height = 4.6cm]{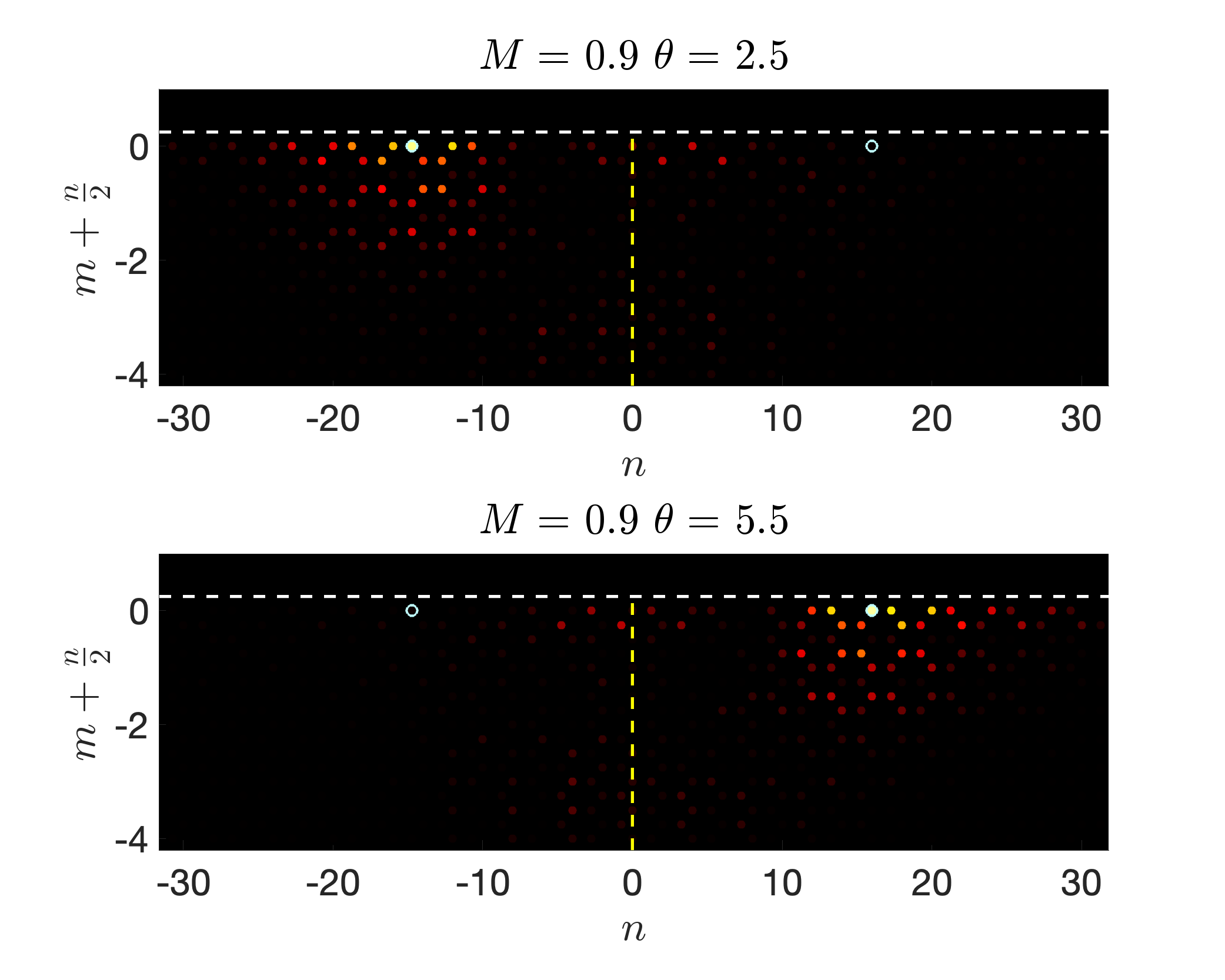}
    \caption{Summary of phase parameters used to control the switch at $\mu = 0$ (see Figure \ref{Fig: SpectrumModes}).
    (Left) The red `hot spot' is where most of the light is traveling along the left lattice edge and the blue `hot spot' is where the most of the light is traveling along the right lattice edge. For a fixed $M=0.9$, the energy of the pulse is direct to the left (Top Right) or to the right (Bottom Right) by adjusting the relative phase, $\theta$, of the initial conditions.} Model parameters are the same as Figure \ref{Fig: SpectrumModes}.
    %\jtc{WHAT ARE YOU SHOWING IN THE FIGURES ON RIGHT?}}
    \label{Fig: Sensor}
\end{figure}

For a chosen value of $\mu$, let the two interface modes be defined as $\rmvec{I}^{(j)}_{nm}(z) = \big[ a^{(j)}_{nm}(z) ~,~ b^{(j)}_{nm}(z)\big]^{\rm{T}}$ for $j = 1,2$. %(see (\ref{edge_soln1}-\ref{edge_soln2})}. 
The superscripts denote the two different interface modes; see %e.g. 
+-markers in Figure \ref{Fig: SpectrumModes}. The initial conditions at $z=0$ are constructed with a ``slowly'' varying envelope, $E_{nm}$, 
in the direction of the interface multiplied by a linear combination of the two interface modes with an adjustable term to control the interference between the modes, written as\\
\begin{equation}
    \begin{bmatrix}
        a_{nm}(0)\\
        b_{nm}(0)
    \end{bmatrix} = E_{nm} \Bigg(
    M\ee^{\ii \theta}     \rmvec{I}^{(1)}_{nm}(0) +\rmvec{I}^{(2)}_{nm}(0)
                    \Bigg) ,
\end{equation}
where we used the envelope $E_{nm} = \mathrm{sech}\left(\nu(m+\frac{n}{2})\right)$, for $\nu$ small. To measure the size of %split between 
the light traveling along the left and right edges, we evaluate the waves at convenient %particular 
lattice sites symmetric about the interface, %on either side of the interface: 
$S_L$ and $S_R$,
at $z= 20$:
\begin{subequations}
    \begin{align}
S_L & = b_{-8,4}(20)\\
S_R & = a_{8,4}(20).
    \end{align}
\end{subequations}
These lattice sites are equidistant from the interface%and 
; at $z=20$ the left and right propagating outputs have essentially %almost completely 
separated. 

In Figure \ref{Fig: Sensor}, we see the results of tuning the linear combination of the two modes. The red and blue `hot spots' represent regions where the light travels almost entirely along the left or right lattice edge, respectively. 
%There is only one region that produces each of the entirely left and right traveling output signals. 
The switching is robust to small changes in the parameters. %This means 
Hence the switching can be achieved while only tuning the relative phase of the two interface modes and keeping the magnitude constant, $M = 0.9$, as seen in Fig. \ref{Fig: Sensor}. Below, we find the optimal magnitude and phase for splitting the pulse.

\begin{figure}
    \centering
    \includegraphics[height = 4.2cm]{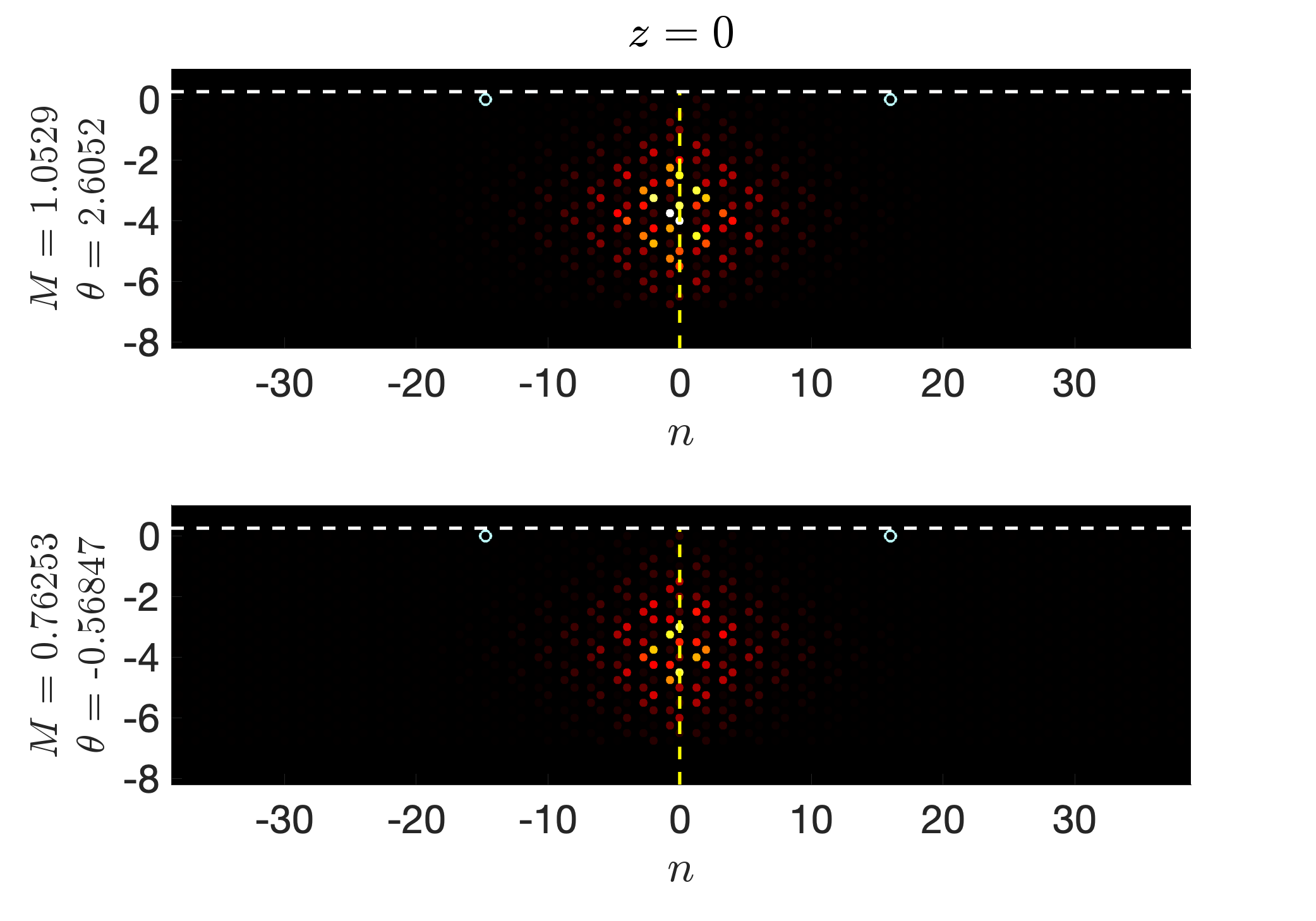}
    \includegraphics[height = 4.2cm]{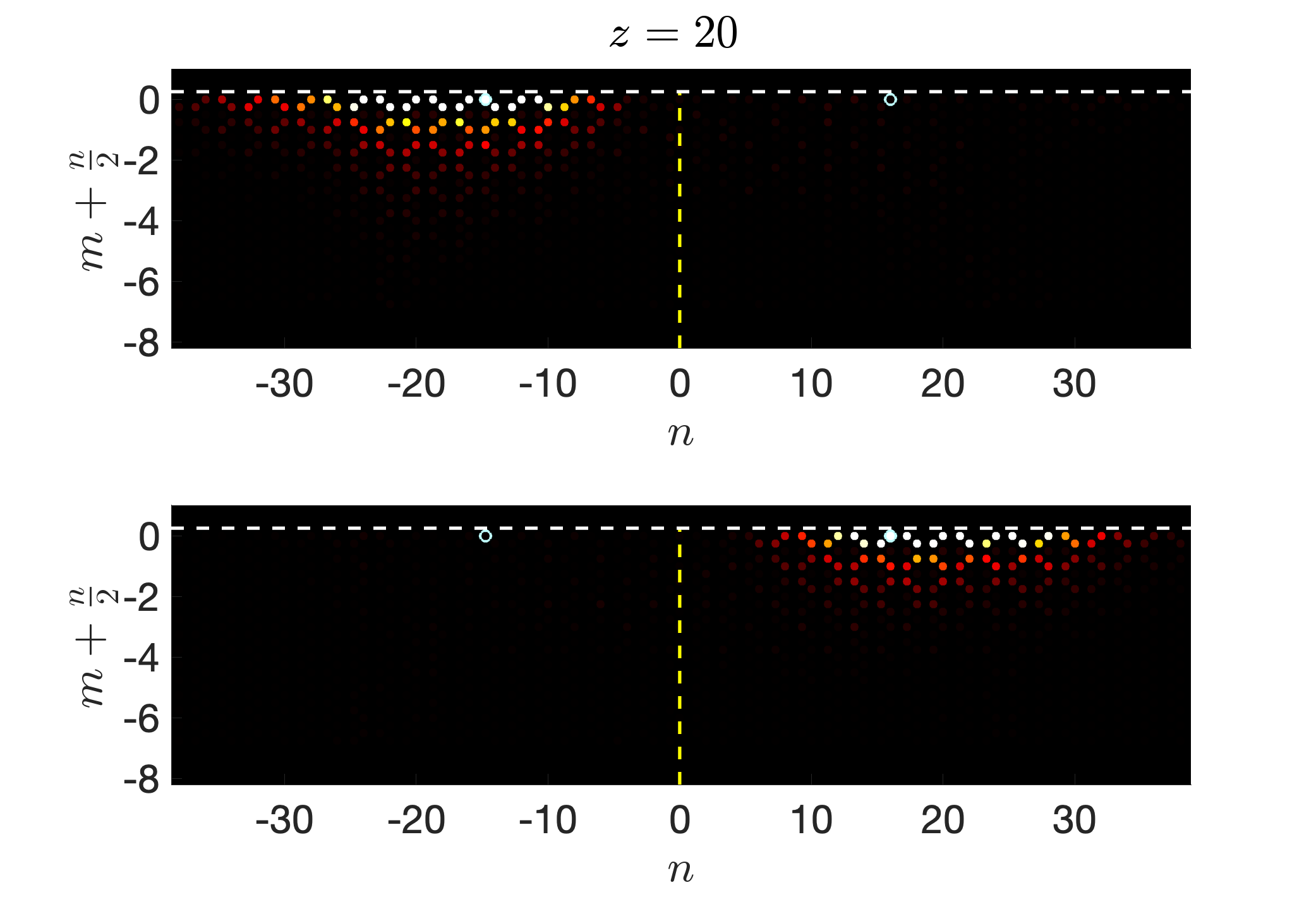}
    \caption{(Left) Initial conditions constructed using the transmission matrix (\ref{Eq: TMatrix}). (Right) Evolution to $z=20$ where we see that the energy has traveled either entirely left (Top) or entirely right (Bottom). Model parameters are the same as Figure \ref{Fig: SpectrumModes}.} 
    \label{Fig: Splitting}
\end{figure}

The switching can be calibrated by sending an unmixed mode of each type, i.e.,
$[a_{nm}^{(j)}(0),\, b_{nm}^{(j)}(0)]^{\rm{T}} = 
E_{nm} \rmvec{I}^{(j)}_{nm}(0)$ for $j= 1,2$, along 
the interface. For measurements that capture the magnitude and phase we define $S_L^{(j)}$ and $S_R^{(j)}$ as the output values on the left and right, respectively, that result from the initial values $E_{nm} I^{(j)}_{mn}(z=0), j=1,2$. The splitting of the light at the edge of the lattice is a linear effect; without loss we can construct a transmission matrix to solve for the initial conditions that will result in wave packets traveling only left or only right. This is because the wave packet being sent into the switch consists of a mixing of the two eigenmodes that exist along the interface as seen in Figure \ref{Fig: SpectrumModes}, while each of the subdomains has a single edge mode that exits along the top boundary.  

Concretely, consider an initial condition $[a_{nm}(0),\, b_{nm}(0)]^{\rm{T}} = E_{nm} \Big( c_1 \rmvec{I}^{(1)}_{nm}(0) + c_2 \rmvec{I}^{(2)}_{nm}(0)\Big)$ which yields output measurements $S_L$ and $S_R$. Mathematically, this system is described by the linear system
\begin{equation}
\label{LinSys}
T\begin{bmatrix}
    c_1 \\ c_2 
\end{bmatrix} = 
\begin{bmatrix}
    S_L \\ S_R 
\end{bmatrix} ,
\end{equation}
where $T$ is the $2\times 2$ transmission matrix which maps coefficients $c_1,c_2$ to output states $S_L,S_R$. To determine %what 
the transmission matrix % is, 
we employ two trials. % runs. 
First, suppose we only excite the first ($j = 1$) interface state, so $c_1=1,c_2 = 0$, and find the first column of the transmission matrix is
%%%%%%%%
\begin{equation}
%T \begin{bmatrix}
%    c_1 \\ c_2 
%\end{bmatrix} = 
T \begin{bmatrix}
    1 \\ 0 
\end{bmatrix}
=
\begin{bmatrix}
    S_L^{(1)} \\ S_R^{(1)} 
\end{bmatrix} .
\end{equation}
%%%%%%%%
Similarly, taking the second interfaces alone ($j = 2$), with $c_1 = 0, c_2 = 1$, we get the second column of the transmission matrix
%%%%%%%%
\begin{equation}
%T \begin{bmatrix}
%    c_1 \\ c_2 
%\end{bmatrix} = 
T \begin{bmatrix}
    0 \\ 1 
\end{bmatrix}
=
\begin{bmatrix}
    S_L^{(2)} \\ S_R^{(2)} 
\end{bmatrix} .
\end{equation}
%%%%%%%%
Hence, the transmission matrix is  given by
\begin{equation}
    T = \begin{bmatrix}
        S_L^{(1)} & S_L^{(2)}\\
        S_R^{(1)} & S_R^{(2)}
    \end{bmatrix}.
    \label{Eq: TMatrix}
\end{equation}

To obtain output states that either completely transmit left or right we use equation (\ref{LinSys}) and invert the transmission matrix. Specifically, to find an interface mode that completely transmits left, and has an output state of $\left[ S_L , S_R \right]^T = \left[ 1,0 \right]^T$, we compute
\begin{equation}
\begin{bmatrix}
    c_{1} \\ c_{2} 
\end{bmatrix}_L = T^{-1}
\begin{bmatrix}
    1 \\ 0 
\end{bmatrix}.
\end{equation}
Similarly, if we want a state to completely transmit to the right, we select $\left[ S_L , S_R \right]^T = \left[ 0,1 \right]^T$ and solve
\begin{equation}
 \quad \mathrm{and} \quad
\begin{bmatrix}
    c_{1} \\ c_{2} 
\end{bmatrix}_{R} = T^{-1}
\begin{bmatrix}
    0 \\ 1 
\end{bmatrix}.
\end{equation}
For all cases considered here, the transmission matrix is nonsingular, so unique solutions are obtained. Using this method, we can solve for the center of the red (left moving) and blue (right moving) regions from figure \ref{Fig: Sensor} as $M = 1.0529$ $\theta= 2.6052$ and $M = 0.76253$ $\theta = 5.7147$, respectively. This is indicated in Figures \ref{Fig: Sensor} and \ref{Fig: Splitting} where the evolution of these initial conditions are shown.

In summary, we use a tight-binding approximation %s 
to find a discrete  system describing two regions of %with 
helically varying (Floquet) waveguides %of 
with opposite driving orientations.  % opposite rotation. %The direction of rotation determines the topological properties of the lattice; e.g. the Chern number. The Chern number of two regions  have opposite signs. 
Along the interface boundary, we find two topologically protected interface modes that propagate unidirectionally.
%We find that interface between the counter rotating lattice regions supports a two dimensional space of topological interface modes and that these modes travel unidirectionally along the interface as slowly varying envelopes of carrier waves. 
We outline the fundamental elements  of a switching mechanism based on the interaction
of the two carrier waves impinging on an edge perpendicular to the direction of the interface.

What distinguishes this switching approach is that it requires no adjustments  to the lattice; all pulse splitting is controlled through the careful preparation (linear combination) of the initial conditions. Said differently, within a fixed lattice system, it is possible to dictate the flow of light; either totally left, totally right, or evenly. Furthermore, for chosen splitting, we have described  a transmission matrix protocol for identifying the optimal initial configuration. That is, one can solve the inverse problem: given a desired output, find  the appropriate input that produces it.
Lastly, we emphasize that the eventual switching can be determined and dictated  in a nonlocal fashion, well away from the switching junction. Intuitively, the initial pulse ``remembers'' its switching instructions for a given lattice configuration.

%The switching relies on the construction of the initial wave input as opposed to manipulation of the lattice itself. 
% This phenomena provides a basis for designing photonic devices based on the propagation of interface modes through lattices with regions of differing topology.  In this paper the two edges are chosen to be perpendicular 
%to each other. But the switching mechanism only relies on the two dimensional space of interface modes  
%directed into two separate edges each admitting only one mode. 
%We expect that many variations of this concept are  
%possible and may also be applicable for other settings involving topological lattices. 

\section{Acknowledgement} 
This project was partially supported by the AFOSR under grants No. FA9550-19-1-0084 and FA9550-23-1-0105.

\appendix
\section{Details}
\label{APP: Details}

The discrete lattice equations (\ref{Eq: DiscreteLatticeEquations}) are derived by tight binding analysis. The ansatz (\ref{Eq: Ansatz}) assumes that the solution to the Schr\"{o}dinger equation (\ref{Eq: NLS}) can be represented by a superposition of orbital functions
defined by (\ref{Eq: OrbitalEquation}) that follow the helical driving of the potential (\ref{Eq: Potential}) with corresponding amplitude functions $a_{nm}(z)$ and $b_{nm}(z)$. After substituting the ansatz (\ref{Eq: Ansatz}) into the Schr\"{o}dinger equation (\ref{Eq: NLS}), we have
\begin{align}
0 = \sum_{(n,m) \in  \mathbb{Z}^2} & \frac{\dd a_{nm}}{\dd z}\phi_{nm}^{1} -\ii a_{nm}\nabla \phi_{nm}^{1} \cdot \rmvec{h}_{n}'(z) +a_{nm}\widetilde{W}_{nm}^{1}\phi_{nm}^{1}  \notag \\
& \frac{\dd b_{nm}}{\dd z}\phi_{nm}^{2} -\ii b_{nm}\nabla \phi_{nm}^{2} \cdot \rmvec{h}_{n+1}'(z) +b_{nm}\widetilde{W}_{nm}^{2}\phi_{nm}^{2} 
\label{Eq: SubAnsatz}
\end{align}
where 
\begin{equation}
\phi_{nm}^{\ell}(z,\rmvec{r}) = \phi\big(\rmvec{r} -\rmvec{r}_{nm}^{\ell} - \rmvec{h}_{n-1+\ell}(z)\big)
\end{equation}
are the orbital functions localized for each lattice site and 
\begin{equation}
\widetilde{W}_{nm}^{\ell}(z,\rmvec{r}) = V(z,\rmvec{r})-\widetilde{V}\big(\rmvec{r} -\rmvec{r}_{nm}^{\ell} - \rmvec{h}_{n-1+\ell}(z)\big)
\end{equation}
is the lattice with the site centered at $\rmvec{r}_{nm}^{\ell}$ removed. Again, $\rmvec{h}_{n}(z) = \rmvec{h}_L(z)$ for $n\leq0$ and $\rmvec{h}_{n}(z) = \rmvec{h}_R(z)$ for $n>0$.

By taking the inner product of (\ref{Eq: SubAnsatz}) with an arbitrary orbital $\phi_{nm}^{\ell}(z,\rmvec{r})$ for each lattice site, we arrive at a system of differential equations for the orbital coefficients %amplitude functions 
\begin{subequations}
\begin{align}
0& = \ii \frac{\dd a_{nm}}{\dd z} + p_{nm}^{(1)}a_{nm}  \\
&\hspace{1cm}+q_{nm}^{(1)} b_{nm} + q_{nm}^{(2)}b_{n-1,m+1}+q_{nm}^{(3)}b_{n-1,m} \\
0& = \ii \frac{\dd b_{nm}}{\dd z} + p_{nm}^{(2)}b_{nm}\\
&\hspace{1cm}+\widetilde{q}_{nm}^{(1)} a_{nm} + \widetilde{q}_{nm}^{(2)}a_{n+1,m-1}+\widetilde{q}_{nm}^{(3)}a_{n+1,m}.
\end{align}
\end{subequations}
The coefficients in the system are 
\begin{subequations}
\begin{align}
    p_{nm}^{\ell} &= \iint \phi_{nm}^{\ell} \big(\rmvec{r}\big)^2 \widetilde{W}_{nm}^{\ell}\big(z,\rmvec{r}\big)~ \dd x~ \dd y\\
    q_{nm}^{(1)} &= \iint \phi_{nm}^{(1)} \big(\rmvec{r}\big)\phi_{nm}^{(2)} \big(\rmvec{r}\big)\widetilde{W}_{nm}^{\ell}\big(z,\rmvec{r}\big)~ \dd x~ \dd y\\
        & \quad  -\ii \iint \phi_{nm}^{(2)}(\rmvec{r}\big) \nabla\phi_{nm}^{(1)}(\rmvec{r}\big)  \cdot \rmvec{h}_{n+1}'(z)~ \dd x~ \dd y\notag\\
    q_{nm}^{(2)} &= \iint \phi_{nm}^{(1)} \big(\rmvec{r}\big)\phi_{n-1,m+1}^{(2)} \big(\rmvec{r}\big)\widetilde{W}_{nm}^{\ell}\big(z,\rmvec{r}\big)~ \dd x~ \dd y\\
            & \quad  -\ii \iint \phi_{n-1,m+1}^{(2)}(\rmvec{r}\big) \nabla\phi_{nm}^{(1)}(\rmvec{r}\big)  \cdot \rmvec{h}_{n+1}'(z)~ \dd x~ \dd y\notag\\
    q_{nm}^{(3)} &= \iint \phi_{nm}^{(1)} \big(\rmvec{r}\big)\phi_{n-1,m}^{(2)} \big(\rmvec{r}\big)\widetilde{W}_{nm}^{\ell}\big(z,\rmvec{r}\big)~ \dd x~ \dd y\\
            & \quad  -\ii \iint \phi_{n-1,m}^{(2)}(\rmvec{r}\big)\nabla\phi_{nm}^{(1)}(\rmvec{r}\big)   \cdot \rmvec{h}_{n+1}'(z)~ \dd x~ \dd y\notag\\
    \widetilde{q}_{nm}^{(1)} &= \iint \phi_{nm}^{(1)} \big(\rmvec{r}\big)\phi_{n-1,m}^{(2)} \big(\rmvec{r}\big)\widetilde{W}_{nm}^{\ell}\big(z,\rmvec{r}\big)~ \dd x~ \dd y\\
            & \quad  -\ii \iint \phi_{n-1,m}^{(1)}(\rmvec{r}\big) \nabla\phi_{nm}^{(2)}(\rmvec{r}\big)  \cdot \rmvec{h}_{n}'(z)~ \dd x~ \dd y\notag\\
    \widetilde{q}_{nm}^{(2)} &= \iint \phi_{nm}^{(1)} \big(\rmvec{r}\big)\phi_{n+1,m-1}^{(2)} \big(\rmvec{r}\big)\widetilde{W}_{nm}^{\ell}\big(z,\rmvec{r}\big)~ \dd x~ \dd y\\
                & \quad  -\ii \iint  \phi_{n+1,m-1}^{(1)}(\rmvec{r}\big) \nabla\phi_{nm}^{(2)}(\rmvec{r}\big)\cdot \rmvec{h}_{n}'(z)~ \dd x~ \dd y\notag\\
    \widetilde{q}_{nm}^{(3)} &= \iint \phi_{nm}^{(1)} \big(\rmvec{r}\big)\phi_{n+1,m}^{(2)} \big(\rmvec{r}\big)\widetilde{W}_{nm}^{\ell}\big(z,\rmvec{r}\big)~ \dd x~ \dd y\\
                & \quad  -\ii \iint \phi_{n+1,m}^{(1)}(\rmvec{r}\big) \nabla\phi_{nm}^{(2)}(\rmvec{r}\big)   \cdot \rmvec{h}_{n}'(z)~ \dd x~ \dd y. \notag
\end{align}
\end{subequations}
These can be reduced by substituting asymptotic approximations for integrals in the $q_{nm}^{j}$'s and $\widetilde{q}_{nm}^{j}$'s coefficients to obtain %at 
\begin{equation}
 q_j(z,\rmvec{w}) = \Big(q_0- \ii s_0 {\bf h}_j'(z) \cdot \rmvec{w} \Big)\ee^{-\sqrt{\mu} \Vert \rmvec{w} \Vert} ,
\end{equation}
where j=L,R and 
\begin{subequations}
    \begin{align}
        q_0\ee^{-\sqrt{\mu} \Vert \rmvec{w}_1 \Vert} &\approx \iint \Big(\phi\big(\rmvec{r}\big)\Big)^2 \widetilde{V}\big(\rmvec{r} - \rmvec{w}_1 \big)~ \dd x~ \dd y\\
        s_0\Vert \rmvec{w}_1 \Vert\ee^{-\sqrt{\mu} \Vert \rmvec{w}_1 \Vert} & \approx \iint \phi_x\big(\rmvec{r}\big) \phi\big(\rmvec{r}-\rmvec{w}_1\big) ~ \dd x~ \dd y.
    \end{align}
\end{subequations}
We arrive at the system of equations given in (\ref{Eq: LeftRightNM1}-\ref{Eq: LeftRightNM2}) and (\ref{Eq: InterfaceNM1}-\ref{Eq: InterfaceNM2}).

\end{document}